\documentclass[12pt,epsf,epsfig]{iopart}
\usepackage{epsfig}
\newcommand \bea{\begin{eqnarray}}
\newcommand \eea{\end{eqnarray}}
\newcommand \beq{\begin{eqnarray}}
\newcommand \eeq{\end{eqnarray}}
\newcommand \ga{\raisebox{-.5ex}{$\stackrel{>}{\sim}$}}

\begin{document}

\title[Bosons and Fermions near Feshbach resonances]{Bosons and Fermions near 
Feshbach resonances}

\author{Henning Heiselberg 
\footnote[3]{Danish Defense Research Establishment (hh@ddre.dk)} 
}

\address{Univ. of Southern Denmark, Campusvej 55, 5230 Odense M, Denmark}

\begin{abstract}

 Near Feshbach resonances, $n|a|^3\gg 1$, systems of Bose and Fermi
particles become strongly interacting/dense. In this unitary limit
both bosons and fermions have very different properties than in a
dilute gas, e.g., the energy per particle approach a value
$\hbar^2n^{2/3}/m$ times an universal many-body constant. Calculations
based upon an approximate Jastrow wave function 
can quantitatively describe recent
measurements of trapped Bose and Fermi atoms near Feshbach resonances.
 
The pairing gap between attractive fermions also scales as
$\Delta\sim\hbar^2n^{2/3}/m$ near Feshbach resonances and is a large
fraction of the Fermi energy - promising for observing BCS
superfluidity in traps. Pairing undergoes several transitions
depending on interaction strength and the number of particles in the
trap and can also be compared to pairing in nuclei.

\end{abstract}




\section{Introduction}

Recent experiments probe systems of bosons
\cite{Claussen} and fermions \cite{Thomas,ENS,Regal,Ketterle,Innsbruck} near
Feshbach resonances by expansion and RF spectroscopy, where
the interactions strengths and densities are large,
$n^{1/3}|a|\gg1$. In the Hamiltonian for a bulk 
system of $N$ fermions or bosons
\bea  \label{Hbulk}
  H &=& \sum_{i=1}^{N} \frac{{\bf p}_i^2}{2m}
  + \sum_{i<j} v({\bf r}_i-{\bf r}_{j}) \,,
\eea 
interacting through two-body potentials $v(r)$, one can then no longer 
apply the dilute limit pseudo-potential approximation: 
$v(r)=4\pi\hbar^2a \delta^3({\bf r})/m$, where $a$ is the scattering length, 

For example, the energy per particle in the dilute limit is
\bea \label{dilute} 
   \frac{E}{N} = 2\pi\hbar^2\frac{an}{m}  \,,\quad n|a|^3\ll 1 \,,
\eea
for bosons and half that for the interaction energy of fermions 
in two spin states. It would lead to
infinitely large positive and negative energies at Feshbach resonances,
$a\to\pm\infty$.
In stead the strongly or dense limit
also known as the unitarity limit is encountered 
where energies (and pairing gaps) are predicted \cite{HH,Cowell}
to approach 
\bea \label{unitarity} 
   \frac{E}{N} =constant\times \hbar^2\frac{n^{2/3}}{m}\ ,\quad n|a|^3\gg 1 \,,
\eea
where the universal {\it constant} is a fundamental many-body parameter 
that only
depends on the spin and the number of spin states of the particle. 
Several calculations and measurements of this constant now exist as 
will discussed in detail below.
On dimensional grounds the energy per particle is expected 
to scale with density as $n^{2/3}$ independent of
the scattering length $|a|\to\infty$. Thus both the Bose and Fermi
interaction energies ``fermionize'' in the unitarity limit.

It is implicitly assumed that both the scattering length and the 
interparticle spacing $r_0=(3/4\pi n)^{1/3}$ are much larger
than the range $R$ of interaction which is usually the case for the cold
atomic clouds. It is also the case for low density neutron matter 
since two neutrons are just unbound
with $^1S_0$ scattering length $a=-18$~fm much larger 
numerically than
the typical $R\simeq1$~fm range of nucleon-nucleon interactions.
In contrast the neutron and proton 
are weakly bound as the deuterium atomic nucleus.

One of the predictions for the novel scaling laws in the unitarity
limit for fermions \cite{HH} and bosons \cite{Cowell} were based on a
calculation using the Jastrow ansatz for the two-body correlations
also referred to as the lowest order constrained variation (LOCV)
approximation \cite{Vijay}.  This is a very useful model as it extends
from both the dilute to the unitarity limit and analytically continues
across Feshbach resonances $a\to\pm\infty$.

Studies of the transition from a Fermi gas of attractive atoms to a
molecular BEC conventionally contain both atom and molecule components
interacting resonantly (see, e.g.,
\cite{Timmermans,Holland,Ohashi,Kohler,Ho,Yurovsky}). The Jastrow
approximation incorporates the strongly interacting limit including
the transformation to molecules very differently. We shall argue that
the transition from a Fermi gas to molecular BEC is smooth in the
sense that two-body correlations gradually build up across the
Feshbach resonance approaching the two-body wave function of molecules
consisting of two bound Fermi atoms.  The large size of the molecules
near Feshbach resonances are therefore included in the two-body
correlation function between atoms.  The transition is continuous as
$a\to\pm\infty$ and will be described within LOCV.

 We shall briefly outline the LOCV approximation and
calculate the energy per particle in detail for both Bose and Fermi
atoms in sections 2 and 3 respectively.  We will show that the results
compares well to recent data on Bosons \cite{Claussen} and Fermions
\cite{Thomas,ENS,Regal} also near Feshbach resonances.
In the remaining sections 4+5 we will discuss BCS pairing in atomic
trap with attractive Fermions and draw a connection to pairing in nuclei
and nuclear matter.

\section{Bosons}

The LOCV method was developed for strongly interacting correlated
fluids as $^4He$, $^3He$ and nuclear matter in \cite{Vijay} and has
more recently been applied to kaon condensation \cite{kaon} and
strongly interacting fermions \cite{HH} and bosons \cite{Cowell}. As
explained in these references the Jastrow wave function 
\bea
\Psi_J({\bf r}_1,...,{\bf r}_N)= \prod_{i<j}f({\bf r}_i-{\bf r}_j) \,,
\eea 
incorporates essential two-body correlations and is a good
approximation for cold dilute and dense bose systems.  We shall extend
this wave function to fermions in the next section.  The pair
correlation function $f(r)$ can be determined variationally by
minimizing the expectation value of the energy, $E/N = \langle \Psi
|H| \Psi \rangle \ /\ \langle \Psi | \Psi \rangle$, which may be
calculated by Monte Carlo methods that are fairly well approximated by
including only two-body clusters.  The basic idea of this method (LOCV)
is that for $r < r_0$ the Jastrow function $f(r)$ approximately obeys the
Schr\"odinger equation for a pair of particles
\bea \label{Schrodinger}
  \left[-\frac{\hbar^2}{m}\frac{d^2}{dr^2} +v(r)\right] rf(r) =\lambda\, rf(r)
   \, ,
\eea
where the eigenvalue energy of two atoms $\lambda=2E/N$.
To take into account many-body effects, which become important
when $r$ is comparable to $r_0$, the conditions
$f(r>d)=1$ and $f^{\prime}(r=d)=0$ are imposed at the healing distance $d$, 
which is determined self consistently from number conservation
\bea \label{sumrule}
  n\int_0^d f^2(r)d^3r = 1 \,.
\eea
In the dilute limit $f(r)\simeq 1$ and so $d=r_0$
whereas in the dense limit $d=(2/3)^{1/3}r_0$. Generally $d\simeq r_0$.

When the range $R$  of the two-body
potential $v(r)$ is small compared to both $|a|$ and $r_0$, 
the boundary condition at short distances is given by the scattering
length $(rf)'/rf=-1/a$ at $r=0$.
Solving the Schr\"odinger
equation gives a wave-function $rf(r)\propto\sin[k(r-b)]$ for positive energies.
The boundary conditions and normalization determine the phase $kb$,
the energy and the healing length $d$.
When positive, the energy $E/N=\hbar^2k^2/2m$ is calculated in terms of the
scattering length (see \cite{Cowell} for details)
\bea 
 \frac{a}{d} = \frac{\kappa^{-1}\tan\kappa -1}{1+\kappa\tan\kappa} 
\,,\label{pos}
\eea
where $\kappa =kd$. 

\vspace{0.cm}
\begin{figure}
\begin{center}
\epsfig{file=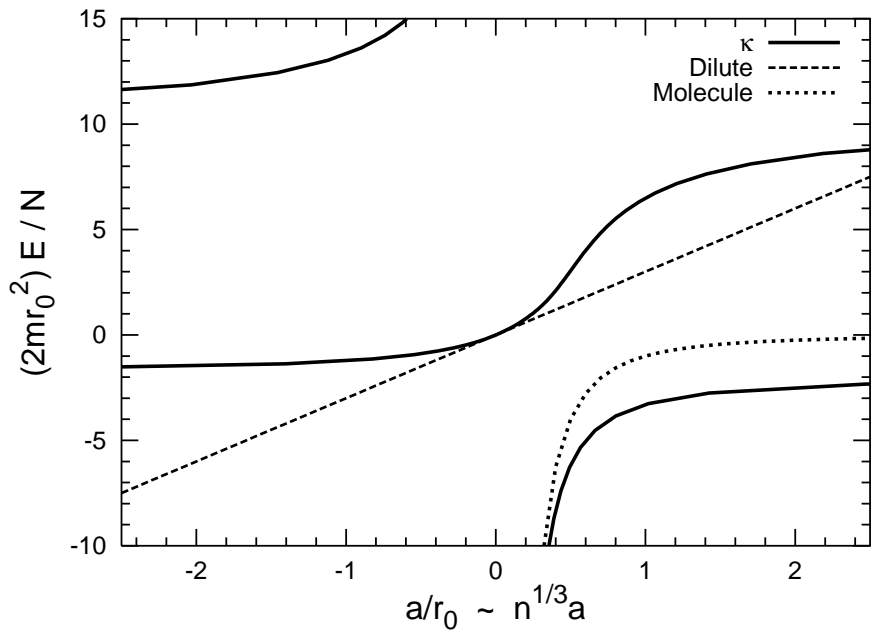,height=12.0cm,angle=-90}
\vspace{.2cm}
\end{center}
\caption{\label{E/N}LOCV calculation of the energy per particle 
$(2mr_0^2)E/N\simeq \pm \kappa^2$ in a BEC vs. scattering length.
Dashed line is the dilute limit result
whereas the dotted line is half
the molecule energy in vacuum $E/N=-\hbar^2/2ma^2$.}
\end{figure}

There is also a negative energy solution as shown in Fig. 1. For
negative scattering lengths such many-boson systems are unstable towards
collapse whereas the many-fermion system is stable as will be discussed below.
Furthermore, the negative energy solution for both bosons and fermions, 
when the scattering length cross over from large and negative to positive,
corresponds to the molecular BEC as we shall see shortly. 
The solution for wave
function to the Schr\"odinger Eq. (\ref{Schrodinger}) 
is $rf(r)\propto\sinh[k(r-b)]$ or
$rf(r)\propto\cosh[k(r-b)]$.  The energy $E/N=-\hbar^2k^2/2m$ is
determined by
\bea 
 \frac{a}{d} = \frac{\kappa^{-1}\tanh\kappa-1}{1-\kappa\tanh\kappa} 
\,.\label{neg}
\eea
The multiple solutions to Eqs. (\ref{neg}) and (\ref{pos})
corresponding to 0,1,2,3... number of
nodes in the (two-body) wave-function $f(r)$ are shown in Fig. (\ref{E/N}).
In the limit $R\to 0$ nodes inside the two-body potential and deeply bound 
states are irrelevant.

In the dilute limit $|a|\ll r_0$ Eqs. (\ref{pos}) and (\ref{neg}) 
give the correct energy per particle, Eq. (\ref{dilute}).
By contrast, in the unitarity limit $|a|\gg r_0$,
Eq. (\ref{pos}) reduces to $\kappa\tan\kappa = -1$
with solutions $\kappa_1=2.798386..$, $\kappa_2=6.1212..$, etc., 
and asymptotically $\kappa_n=n\pi$ for integer $n$. 
The negative energy solution
to Eq. (\ref{neg}) reduces in the unitarity limit to
$\kappa\tanh\kappa = 1$  for $a\gg r_0$ with solution $\kappa_0=1.1997...$
As the scattering length cross over from $-\infty$ to $+\infty$ the negative
energy state is analytically continued towards the molecular bound state
with $E/N=-\hbar^2/2ma^2$ as $a\to +0$.
Therefore the LOCV approximation to many-body systems
correctly reproduces the energies in both the dilute limit and the molecular
limit. In the unitarity limit it correctly predicts qualitatively the $n^{2/3}$
scaling of particle energies of Eq. (\ref{unitarity}). 
Quantitative comparison to experimental data
for cold bose and fermi atoms in the unitarity limits will be given below.

 The energy of a repulsive BEC gas as it approaches a Feshbach resonance,
$a\to +\infty$, is of special interest
\bea
 \frac{E_1}{N} = \left(\frac{3}{2}\right)^{2/3}
\frac{ \hbar^2\kappa_1^2}{2m r_0^2} 
    = 13.33 \hbar^2\frac{n^{2/3}}{m}\ . 
\label{hde}
\eea

\vspace{0.cm}
\begin{figure}
\begin{center}
\epsfig{file=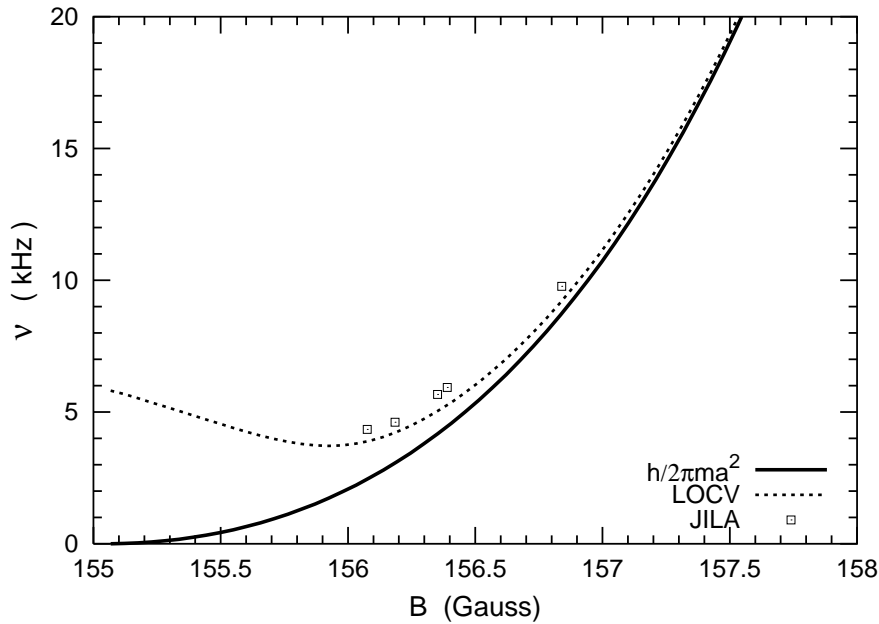,height=12.0cm,angle=-90}
\vspace{.2cm}
\end{center}
\caption{\label{JILA.ps}Frequency of the atomic-molecular transition
in a $^{85}$Rb BEC as function of magnetic field near the Feshbach resonance.
The full curve is the in vacuum transition frequency for two atoms to a 
molecule $\nu=\hbar/2\pi ma^2$. The LOCV prediction of Eq. (\ref{nu})
for the transition of two atoms to a molecule in the many-body 
system is shown with dashed curve. Data from JILA \cite{Claussen}.}
\end{figure}

These energies can be compared to the JILA experiment \cite{Claussen} 
where the scattering length is tuned near the
Feshbach resonance by a magnetic field $B$ as
\bea
   a(B) = a_{bg}\left( 1-\frac{\Delta}{B-B_{Feshbach}}\right) \,.
\eea
Coherent atom-molecule oscillations are induced in the BEC
at a frequency $\nu(B)$.
We take this frequency as the difference between
the mean field energy of two atoms in the
atomic state and the molecular state
\bea \label{nu}
  \nu = \frac{2}{h} \frac{E_+-E_-}{N}
  = \frac{\hbar}{2\pi md^2}\left(\kappa_+^2-\kappa_-^2\right) \,.
\eea
Here, the indices + and - refer to the (lowest) positive and negative energy
branches of Fig. (1) respectively 
as function of a. The negative energy branch approaches
the molecular energy, i.e. $E_-/N\simeq-\hbar^2/2ma^2$ as $a\to +0$.
The resulting frequency is shown in Fig. (\ref{JILA.ps}). 

At the Feshbach resonance
the LOCV frequencies approaches a constant value 
which according to Eqs. (\ref{nu})
is $\nu=\hbar(\kappa_1^2+\kappa_0^2)/2\pi md^2=2.51\hbar n^{2/3}/m$
for the transition of two atoms in the unitarity limit to a molecular state.
For the $^{85}Rb$ BEC of average density 
$n=2\times 10^{-12}cm^{-3}$ \cite{Claussen} this results in a frequency
$\nu=5.9$kHz at the Feshbach resonance 
as seen in Fig. (\ref{JILA.ps}). The calculated transition frequency
is in nice agreement with the JILA data close to the resonance. 
The calculated frequencies have 
a minimum just above the resonance of order $\sim 4$kHz because the atomic
energy decrease faster with detuning (i.e., decreasing scattering
length) than the molecular energy. 

 The absence of experimental data in the 155-156G region is due to
overdamping of the atom-molecule oscillations. The imaginary part of
the self-energies may in this region be so large near the Feshbach
resonance that the quasi-particles are not well defined.

The effective range has been neglected in our calculations because we
have assumed that the range of interaction is small as compared to
scattering lengths and interparticle distances.  The effective range
of the interaction is by definition equal to the range for a square
well potential. However, when the two-body interaction has a long tail
and the scattering length is not large, then corrections to the
scattering amplitude and particle energies from the effective range
appear. This is the case in Fig. (\ref{JILA.ps}) off the resonance
where the scattering length becomes small.
Around $B\ga 157$G the frequency is slightly increased \cite{Duine}
improving the agreement with data.

The depletion of the condensate can also be calculated 
in the LOCV approximation from $f(r)$. The condensate fraction is \cite{Cowell}
\bea \label{CF}
  \frac{n_0}{n} = 1-n\int \left[1-f(r)\right]^2 d^3r 
   = \left(\frac{d}{r_0} \right)^3 \left[ \frac{6}{\kappa^3} 
(\sin \kappa -\kappa \cos \kappa) -1 \right] \,,
\eea
which predicts quenching of the BEC for $na^3\ga 0.6$.
The condensate fraction differs from the dilute limit fraction:
$n_0/n =1-(4/\sqrt{3}\pi)(a/r_0)^{3/2}$,
as well as from the hard-sphere potential and exact quantum Monte
Carlo calculations.

The three-boson problem cannot generally be described by the
scattering length only but depends on a three-body parameter as well
\cite{Braaten} which is referred to as non-universality.  It was shown
in \cite{Jonsell}, however, that for three bosons in a trap the energy
was universal in both the dilute and in the unitarity limit - but not
between these limits. Thus we may expect that the many-boson problem is
universal as well in these two limits. The three-fermion problem is
simpler in the sense that the Pauli exclusion principle prohibit
Efimov states and therefore no three-body parameter enters, and the
three- and many-body problem is universal.

\section{Fermions}

In the unitarity limit, $n|a|^3\gg1$, a system of fermions 
is dense and/or strongly interacting. 
The energy per particle
has been calculated within Galitskii resummation, the LOCV method and
recently by fixed node Green's function Monte Carlo (FN-GFMC).
They find that the interaction energy does not extrapolate to $\pm\infty$ as
the dilute result of Eq. (\ref{dilute}) does, 
but approaches a universal constant times
the kinetic or Fermi energy, $E_F=\hbar^2k_F^2/2m$, as in Eq. (\ref{unitarity})
\bea\label{Dense}
   \frac{E}{N} &=& E_{kin}+E_{int} =  \frac{3}{5} E_F[1+\beta] \,.
\eea
Here, $\beta=E_{int}/E_{kin}$ is a universal many-body parameter in the unitarity
limit, which only depends on the
number of spin states $s$ through the Pauli principle and the density
$n=sk_F^3/6\pi$.
For two spin states $\beta$ is
displayed in Table 1 for the various calculations and recent experiments
as will be described in the following subsections. 
Recent measurements \cite{Thomas,ENS,Regal,Ketterle,Innsbruck} of $\beta$
seem to confirm this unitarity limit near Feshbach resonances.

\subsection{Galitskii resummation}
 
The scattering amplitude, which in the dilute limit is simply $f=-a$,
can in the interacting many-body system be expanded in powers of
scattering length via Galitskii resummation of ladder diagrams
\cite{Galitskii}.  It was shown in \cite{HH} that resummation of
scattering amplitudes lead to a $f\sim k_F^{-1}$ scaling of the real
part of the scattering amplitude in the unitarity limit.
Brueckner, Bethe
and Goldstone \cite{BBG} pioneered such studies for nuclear matter and
$^3He$, which are more complicated cases 
since the range of interactions, scattering lengths and
repulsive cores all are comparable in magnitude. 
 In our case the
range $R$ of interaction is small, $k_FR\ll1$, and therefore all
particle-hole diagrams are negligible.
This is opposite to electro-magnetic interactions where the strength of the
interaction is small but the
range of interactions is large and must be screened by Debye or Landau damping
implicit in loop diagrams.
In the Galitskii resummation all
higher order particle-particle
and hole-hole diagrams are found to contribute to the same order
when $k_F|a|\gg1$. 
Due to the very restricted phase space such higher
order terms are usually neglected as in standard Brueckner theory.
Truncating the resummation to second order one finds for two spin states
\cite{HH} that $\beta=-175/27(11-2\ln2)=-0.67$ in the unitarity limit.

\begin{table}
\caption{\label{table}Calculations and measurements of the ratio of 
interaction to kinetic energy
 $\beta=E_{int}/E_{kin}$  in the dense/strongly interacting regime
$a\to -\infty$ for
two spin states.}
\begin{center}
\begin{tabular}{lll}
\br
-$\beta$ & Calc./Exp. & Reference \\
\mr
0.26$\pm$0.07 & Duke exp. &\cite{Thomas}\\
0.3-0.4 & ENS exp. &\cite{ENS}\\
0.4-0.6 & JILA exp. &\cite{Regal}\\
0.68$\pm$0.1 & Innsbruck exp. &\cite{Innsbruck}\\
0.67 & Galitskii approx. & \cite{HH}\\
0.54 & LOCV approx. &\cite{HH}\\
0.56$\pm$0.01 & FN-GFMC &\cite{Carlson}\\
\br
\end{tabular}
\end{center}
\end{table}

\subsection{LOCV}

The unitarity limit can also be studied in the LOCV approximation.
For fermions the many-body wave function 
must be anti-symmetric. To first approximation one
includes a Slater wave function $(\Phi)$, 
which is the ground state of free fermions
\bea
\Psi_{JS}({\bf r}_1,...,{\bf r}_N)= 
\Phi\prod_{i<j'}f({\bf r}_i-{\bf r}_{j'}) \,,
\eea 
We refer to Ref. \cite{Carlson} where also a BCS wave function including 
pairing is employed.
Because $\Phi$ insures that same spins are spatially anti-symmetric, 
the Jastrow wave function only applies to particles with different spins
(indicated by the primes).
Therefore, the number conservation also applies to
different spin states only and Eq. (\ref{sumrule}) should be multiplied
by a factor 
$(s-1)/s$ on the left hand side which affects the healing length.
For example, in the unitarity limit $a\to -\infty$ 
the healing length now approaches
$d=r_0(2s/(s-1)3)^{1/3}=(3\pi/(s-1))^{1/3}k_F^{-1}$.

Otherwise the LOCV calculation for fermions follows the lines as for bosons
and the interaction energy is $E_{int}=\kappa^2/2md^2$ as calculated above. 
In addition the kinetic energy $(3/5)E_F$ appears due to the Slater
ground state. The total energy becomes
\bea \label{LOCV-F}
  \frac{E}{N}&=&\frac{3}{5}E_F \pm
   \frac{\hbar^2\kappa^2}{2md^2} \,,
\eea
where the $\pm$ refers to the positive and negative energy states 
discussed above. The energy becomes \cite{error}
\bea \label{ELOCVF}
    \frac{E}{N} &=&  \frac{3}{5} E_F  - \frac{\hbar^2\kappa_0^2}{2md^2}
     =  E_F \left[ \frac{3}{5}
   - \left(\frac{s-1}{3\pi}\right)^{2/3} \kappa_0^2  \right]  \,.
   \label{eWS}
\eea
With $\kappa_0=1.1997..$ we 
thus obtain in the unitarity limit $\beta=0,-0.54, -0.85,-1.12,....$, for 
$s=1,2,3,4,...$ respectively.

 For the gas to be stable towards collapse the energy must be
positive, i.e.,  $1+\beta>0$. Therefore the LOCV approximation
predicts that up to $s\le3$ spin states are stable in the unitarity limit
whereas in the
Galitskii approximation only $s=1,2$ are stable.  The marginal
case $s=3$ may be studied with $^6$Li atoms since they have three
spin states with broad and close lying Feshbach
resonances \cite{Ketterle}. 
The stability of two spin states
towards collapse has recently been confirmed for a $^{6}$Li and
$^{40}$K gases near
Feshbach resonances. It has long been known that neutron star matter
\cite{NS} with two spin states likewise has positive energy at all densities
whereas for symmetric nuclear matter with two spin and two isospin
states, i.e.  $s=4$, the energy per particle is negative. Nuclear matter
is therefore unstable
towards collapse and subsequent implosion, spinodal decomposition and
fragmentation at subnuclear densities\cite{HPR}. 
Above nuclear saturation densities,
$k_FR\ga1$, short range repulsion stabilizes matter up to maximum masses
of neutron stars $\sim 2.2M_\odot$, where gravitation makes such stars
unstable towards collapse.

\subsection{Fixed-node Green's function Monte Carlo (FN-GFMC)}

In Ref. \cite{Carlson} the energy of a finite number of Fermions in a
box has been calculated within FN-GFMC at a
Feshbach resonance. Since the gas is in a meta-stable state with
typical $\sim 40$ lower lying molecular states, the number of nodes in
the Jastrow two-body wave-function must be fixed in Monte Carlo 
calculations. At
the Feshbach resonance $a\to -\infty$
they find that the energy of the system increase with the
number of particles as $E=(1+\beta)NE_{kin}$ with
$\beta=-0.56\pm0.01$.  This is an upper limit
since FN-GFMC is also a variational calculation of the energy.

Pairing favors an even number of particles and from the odd-even
staggering of $E(N)$ a pairing gap $\Delta\simeq0.54E_F$ is extracted
from the FN-GFMC calculations at the Feshbach resonance.

When comparing the various calculations of $\beta$ it should be noted
that the FN-GFMC value $\beta=-0.56$ includes pairing energies whereas
the LOCV and the Galitskii approximations did not. Excluding
pairing the FN-GFMC leads to a higher energy state with $\beta=-0.44$
\cite{Carlson}. In comparison the expected contribution
$(3/8)\Delta^2/E_F$ from pairing energies in the dilute limit would
lead to an even larger correction if the pairing gap in the unitarity
limit is inserted.

\vspace{0.cm}
\begin{figure}
\begin{center}
\epsfig{file=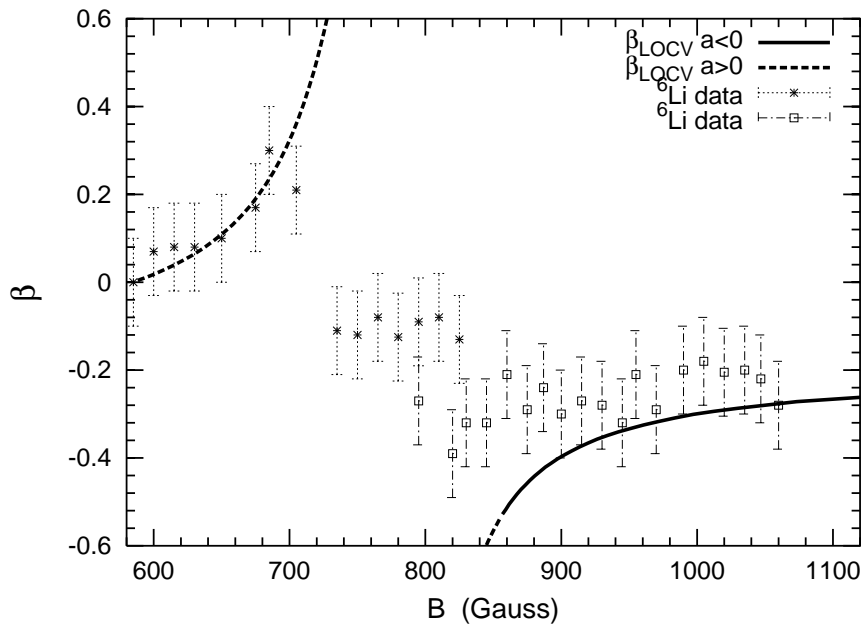,height=12.0cm,angle=-90}
\vspace{.2cm}
\end{center}
\caption{\label{ENS.ps}
Ratio $\beta$ of the interaction energy over the kinetic energy
vs. magnetic field for $^{6}$Li.
Predictions from Eq. (\ref{LOCV-F}) with $\kappa$ calculated within
LOCV are show with full and dashed curves for negative and positive
scattering length respectively. The data, density and 
scattering length vs. magnetic
field near
the Feshbach resonance at 855G is taken from \cite{ENS}.}
\end{figure}

\subsection{Experiments}

Several experiments with trapped Fermi atoms have recently measured energies
in the strongly interacting or dense limit near Feshbach resonances.
The energy in the trap (excluding that from the harmonic oscillator potential)
is $E/N=(3/8)E_F\sqrt{1+\beta}$ where $E_F=(3N)^{1/3}\hbar\omega$ 
is the Fermi energy in a trapped non-interacting gas.

The Duke group \cite{Thomas} measured the energy of $^6$Li Fermi atoms 
near a Feshbach resonance from expansion energies. 
After correcting for thermal energies, they find $\beta=-0.26\pm0.01$
close to the Feshbach resonance $k_Fa\simeq -7.4$.

Bourdell et al. at ENS \cite{ENS} measure expansion energies for
$^6$Li near a Feshbach resonance. They find $\beta=-0.4\pm0.1$ as
$a\to -\infty$ as shown in Fig. (\ref{ENS.ps}).  The situation is more
complicated at the other side of the resonance for positive $a$, where
the ratio increases up to $\beta=0.2\pm0.1$ in agreement with LOCV
predictions, but then drops to a plateau close to the resonance.  The
LOCV calculations compare well to ENS data as $a\to-\infty$ whereas
$\beta(a\to+\infty)$ exceeds the data especially at the plateau.
Whether this is due to molecule formation \cite{Kokkelmans} or finite
temperature, which can have a strong effect on $\beta$ \cite{Muller},
should be investigated.

 The MIT experiment \cite{Ketterle} on $^6$Li and the JILA experiment
on $^{40}$K \cite{Regal} measure the transition frequency between
three hyperspin states which confirm the unitarity limitation to
energies and frequencies. The JILA experiment is performed at two
densities and does not find a plateau. In extracting the JILA value
for $\beta$ as given in Table 1, the reduction factor $(s-1)$ in
Eq. (\ref{ELOCVF}) is replaced by $s$ because of the specific
population of spin states in their RF experiment.

 With the discovery of a molecular BEC the Innsbruck group
\cite{Innsbruck} has been able to measure the size of the atomic
cloud, which scales with $(1+\beta)^{1/4}$, around the Feshbach
resonance at very low temperatures. They find $\beta=-0.68\pm0.1$
compatible with the theoretical estimates.

\section{Pairing of Fermi atoms in harmonic oscillator traps}

BCS pairing and superfluidity is expected for trapped  Fermi atoms 
with attractive interactions. 
As atomic gases have widely tunable number of particles,
densities, interaction strengths, temperatures, spin states, and other
parameters, they hold great promise for a more general understanding
of pairing phenomena in solids, metallic clusters, grains, nuclei,
neutron stars and quark matter.

In a uniform dilute 
gas with sufficiently strong interactions or large number of particles
the bulk limit is reached, and the pairing gap is at zero temperature
\cite{Gorkov,gap}
\bea
   \Delta = E_F \frac{8}{e^2} (4e)^{s/3-1} 
   \exp\left[\frac{\pi}{2ak_F}\right] \, .  \label{Gorkov}
\eea
In the unitarity limit, $k_F|a|\ga 1$, the pairing gap approaches a
finite fraction of the Fermi energy \cite{HH} just as
the energy per particle. The FN-GFMC 
calculations discussed above \cite{Carlson} find that the odd-even
staggering energy or pairing gap in bulk is $\Delta\simeq 0.54E_F$
which is quite close the value $\Delta\simeq 0.49E_F$ obtained by
extrapolating Eq. (\ref{Gorkov}) to $a\to -\infty$ for two spin states.

Near a Feshbach resonance the strongly interacting Fermi gas becomes
unstable towards molecule formation and 
the critical temperature
$T_c=(\gamma/\pi)\Delta\simeq0.567\Delta$
for BCS superfluidity is
expected to cross-over towards the slightly smaller critical temperature
for forming a Bose-Einstein condensate of molecules \cite{Randeria,SHSH}.
If we take $\Delta=0.54E_F$ from FN-GFMC
both the BCS and BEC critical temperatures, $T_c\simeq 0.5\Delta$,
are therefore around or above the
lowest temperatures reported for trapped Fermi atoms
\cite{Thomas,ENS,Regal,Ketterle,Innsbruck}. This bodes well for
observing BCS superfluidity in atomic traps and
establishing the unitarity scaling of pairing gaps.

\vspace{0.cm}
\begin{figure}
\begin{center}
\psfig{file=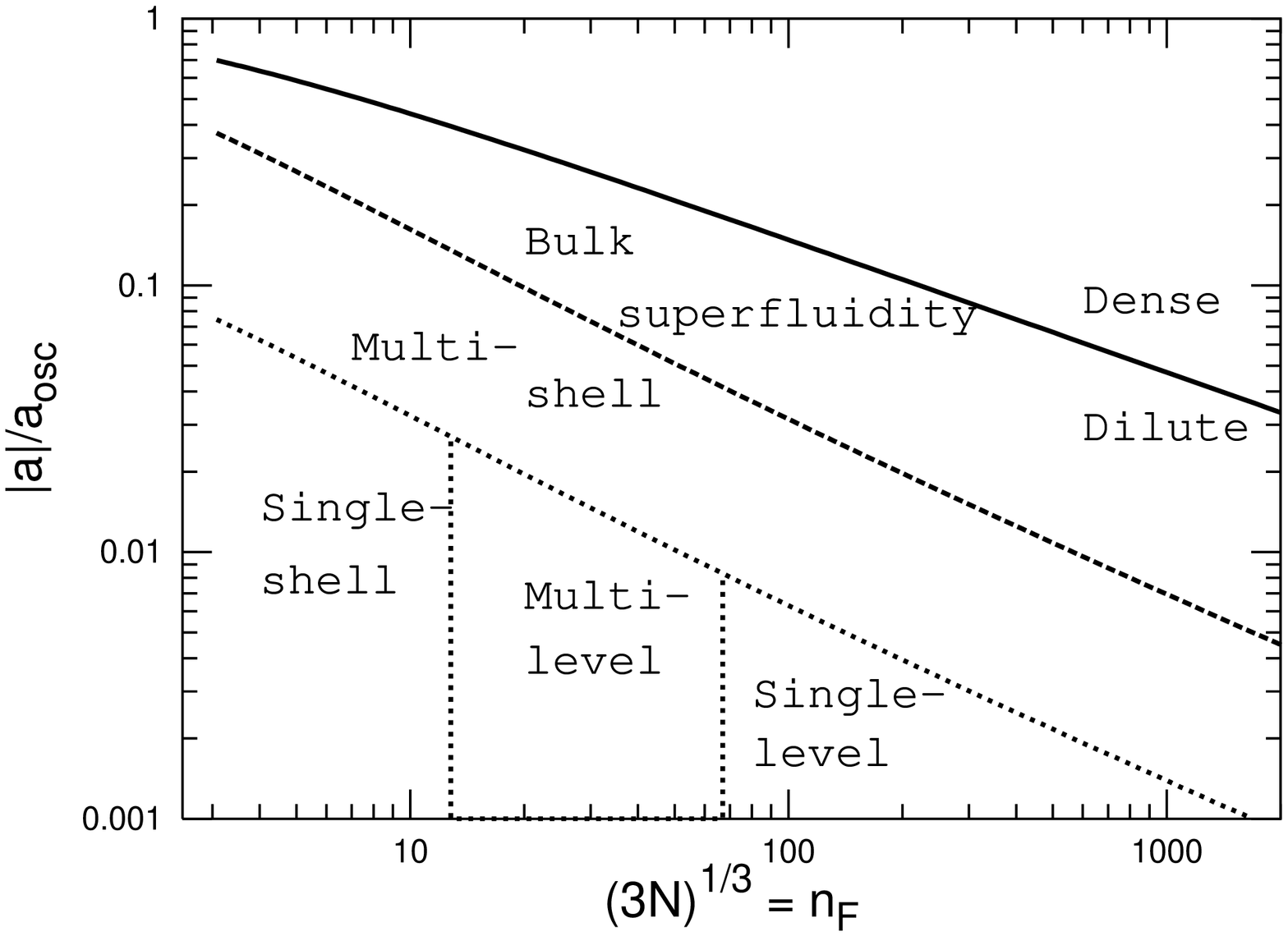,height=11.0cm,angle=-90}
\vspace{.2cm}
\begin{caption}
{Diagram displaying the regimes for the various pairing mechanisms
(see text) at zero temperature
in h.o. traps vs. the number of particles $N=n_F^3/3$
and the interaction strength $a<0$.
The dotted lines indicate the transitions between single-shell pairing 
$\Delta=G$, multi-level, single-level, and multi-shell pairing.
At the dashed line determined by $2G\ln(\gamma n_F)=\hbar\omega$ the pairing
gap is $\Delta\simeq\hbar\omega$, and it 
marks the transition from multi-shell pairing to bulk superfluidity
Eq. (\ref{Gorkov}). The pairing gap is $\Delta=0.54E_F$
above the full line $n|a|^3\ge 1$, which separates the dilute from 
the dense gas. From \cite{A}
}
\end{caption}
\end{center}
\label{phase}
\end{figure}

 As cooling improves further and smaller pairing gaps may be found,
new pairing regimes take over that are important in other systems as
solids, nuclei, neutron stars, etc.  We shall outline the various
pairing regimes shown in Fig. (\ref{phase}) vs. density interaction
strength and relate it to BCS pairing in bulk.

The various pairing regimes shown in
Fig.  (\ref{phase}) can, except from the dense or unitarity limit,
be calculated from in the dilute limit Hamiltonian
\bea  \label{H}
  H &=& \sum_{i=1}^{N} \left( \frac{{\bf p}_i^2}{2m} +
    \frac{1}{2} m\omega^2 {\bf r}_i^2 \right)
  + 4\pi\hbar^2\frac{a}{m} \sum_{i<j} 
     \delta^3({\bf r}_i-{\bf r}_{j}) \,,
\eea 
for $N$ atoms in a spherical harmonic oscillator (h.o.) potential.
When energies are measured in units of $\hbar\omega$, and lengths in 
$a_{osc}=\sqrt{\hbar/m\omega}$ only two parameters remain in the
Hamiltonian, namely the number of particles and the
interaction strength as plotted in  Fig. (\ref{phase})

When the traps contain relatively few atoms that are weakly
interacting (lower left corner of Fig. (\ref{phase})) 
the level spectra are highly degenerate due to the SU(3)
symmetry of the spherical h.o. potential. Due to the Pauli
principle the h.o. shells $n=0,1,...,n_F$ are filled up.
Only the top Fermi shell $n_F\simeq(3N)^{1/3}$, where $N$ is the number of
Fermi atoms, may be partially filled. The levels with angular momentum
$l=n_F,n_F-2,....,1$ or 0 are all degenerate in the weakly interacting limit
due to SU(3) symmetry.
Consequently, pairing is enhanced as the gap generally increases 
with the number of states,
and leads to a supergap \cite{HM}
\bea \label{G}
 G = \frac{32\sqrt{2n_F}}{15\pi^2} \frac{|a|}{a_{osc}}\hbar\omega \,.
\eea
For stronger interactions multi-shell pairing also takes place
whereas for more particles the stronger mean field cause 
level splitting, which reduce pairing towards single
level pairing \cite{HM,BH,A} and shows up as a distinct shell structure
with h.o. magic numbers.

Varying the number of particles and interaction strengths thus allows us to
investigate a wide range of pairing mechanisms that apply to many other
physical systems, e.g., nuclei and neutron stars as will be discussed in
the following section.

\section{Pairing in nuclei and nuclear matter}

The nuclear mean field is often approximated by a simple h.o. form and
the residual effective pairing interaction by a delta force in order
to obtain some qualitative insight into single particle levels,
pairing, collective motion, etc., (see, e.g., \cite{BM,FW}).  We can
therefore compare pairing in nuclei to that in traps as
investigated above, once the h.o. potential is adjusted to describe nuclei.

\begin{figure}
\begin{center}
\psfig{file=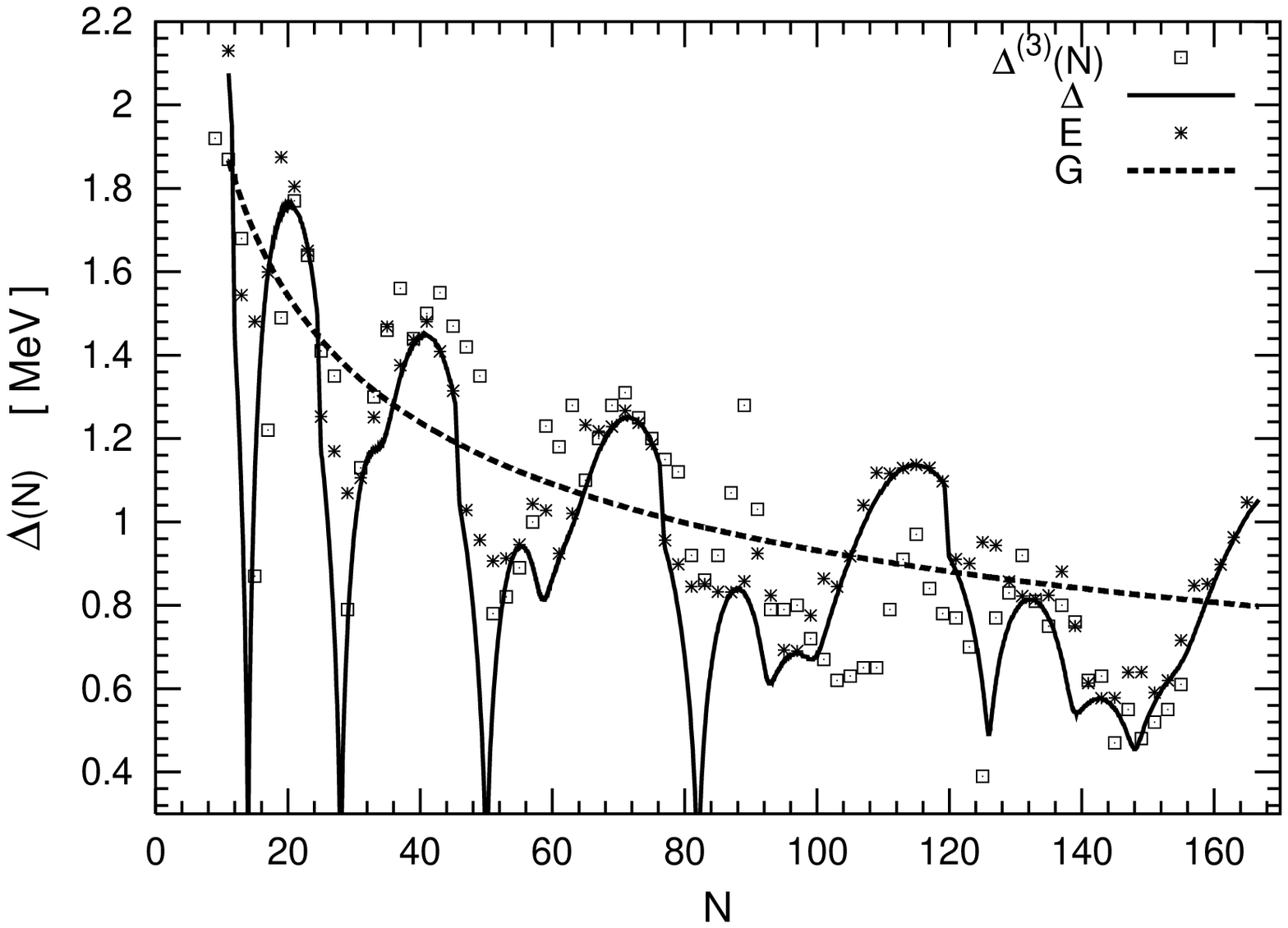,height=11.0cm,angle=-90}
\vspace{.2cm}
\begin{caption}
{Neutron pairing $\Delta^{(3)}(N)$ 
vs. the number of neutrons. The experimental values have been averaged over
isotopes \protect\cite{Audi}. The calculated gaps $\Delta$ and
quasi-particle energies $E$ are obtained
from the gap equation (see text) with effective coupling strength 
$a=-0.41$~fm. The supergap $G$ is shown with dashed line. From \cite{A}.}
\end{caption}
\end{center}
\label{N}
\end{figure}

As described in \cite{A} the anharmonic nuclear mean field is stronger
and of opposite sign than that in atomic traps. It also contains a
strong spin-orbit force which change the magic numbers from the
h.o. shells. Correcting for that the quasi-particle energies and pairing
gaps were calculated by solving the Bogoliubov-deGennes equations which
can be expressed as gap equation. The pairing depends only on one parameter,
namely the effective scattering length in the Hamiltonian $a=-0.41$fm 
that was obtained from fitting the
pairing gaps to the odd-even staggering binding energies of 
neutrons and protons in nuclei (see Fig. \ref{N}).
The shell structure and the average pairing is well reproduced.
If nuclei were to be placed in the pairing phase diagram of Fig.
(\ref{phase}) they would lie in the transition region between
the single-shell, multi-shell and single-level pairing regions.
Because the multi-shell increase pairing and single-level decrease
pairing the average gaps are to a good approximation given by the supergap of
Eq. (\ref{G}). Adjusting the h.o. frequency and oscillator length to
that in nuclei with constant central densities of $n=0.15$fm$^{-3}$
the supergap becomes
\bea
  \Delta\simeq G= \frac{|a|}{0.41{\rm fm}} \frac{5.5{\rm MeV}}{A^{1/3}} 
  \,.
\eea
The supergap predicts a $A^{-1/3}$ scaling in good 
agreement with nuclear pairing data. It differs slightly from the standard
$A^{-1/2}$ scaling in the Bethe-Weisz\" acker
liquid-drop formula. 

The supergap is robust in the sense that it does not depend on the
level spectra or other details, and thus allows us to extract the
effective scattering length directly.

For very large nuclei multi-shell pairing becomes increasingly
important and pairing approaches that in bulk matter. With the
effective nucleon coupling constant $a=-0.41$fm extracted from pairing
gaps in finite nuclei, we can estimate the $^1S_0$ pairing of both
neutrons and protons in nuclear matter from Eq. (\ref{Gorkov})
\bea
  \Delta \simeq 1.1 {\rm MeV} \,.
\eea

Neutron star matter has a wide range of densities and is very
asymmetric, $Z/A\simeq 0.1$. One can attempt to estimate of the pairing gaps as
function of density from the gap in bulk, Eq. (\ref{Gorkov}), with
$a\simeq -0.41$~fm.
However, the effective interaction $a$ is
density dependent. At higher densities we expect the effective interaction to
become repulsive as is the case for the nuclear
mean field at a few times nuclear saturation
density. At lower densities the effective scattering length should approach
that in vacuum which for neutron-neutron scattering
is $a(^1S_0)\simeq -18$~fm.

Neutron stars are so cold that neutron and proton superfluidity is expected
in most of the surface layers. The depinning of superfluid vortices
can explain neutron star glitches in star quakes \cite{Pines}.

\section{Summary and Outlook}

We list the main results discussed above and add a few topics
for future investigation:

\begin{itemize}

\item In the unitarity limit near Feshbach resonances, $n|a|^3\gg 1$,
where bosons and fermions interact strongly or are dense, the particle
energies (and pairing gaps) scale like $\hbar^2n^{2/3}/m$ times an
universal constant.

\item
For $^{85}$Rb bosons recent experiments \cite{Claussen} confirm
the unitarity limit and measure a transition frequency $\nu\simeq 5$kHz
near the Feshbach resonance in agreement with the predictions from
LOCV.
Experiments at other densities are underway to check the ``fermion'' $n^{2/3}$
dependence of the energy per particle.

\item The unitarity limit has also been confirmed for Fermi atoms near 
Feshbach resonances.
Measurements of the universal parameter $\beta=E_{int}/E_{kin}\simeq -0.5$ as
$a\to-\infty$ are compatible with theoretical predictions.

\item The pairing gap in traps with attracting Fermi atoms is $\Delta\simeq
0.5E_F$ near Feshbach resonances and the critical temperature
$T_c\simeq 0.5\Delta$. Thus superfluidity
may already have been achieved in recent experiments.

\item Cooling further will eventually make measurements of smaller pairing gaps
possible.  By
varying the number of particles and interaction strengths the pairing
in h.o. traps with attractive Fermi
atoms undergoes several pairing phases. These exhibit a wide range of pairing
mechanisms that apply to many other physical systems.

\item Pairing of neutrons and protons in nuclei is similar to
super-pairing in atomic traps. The supergap scales as
$\Delta\simeq5.5$~MeV$/A^{1/3}$ with mass number. Nuclei also extend into
the multi-level pairing regime seen as a strong shell dependence with
magic numbers.

\item
Large nuclei approach  the $^1S_0$ pairing bulk 
in nuclear matter $\Delta\simeq 1.1$MeV for both neutrons and protons.
Pairing will, however, vary with density and asymmetry in neutron star matter.
Low neutron densities is also dominated by a large negative scattering
length $a(^1S_0)\simeq-18$fm due to a Feshbach resonance in the NN channel.

\item
Mixing fermionic with bosonic atoms allow sympathetic 
cooling \cite{Ketterle,Stoof,Hulet,Inguscio,Salomon} 
and thus to study weaker pairing.
Induced interactions between fermions and bosons generally
enhance pairing \cite{gap}.

\item
A systems of fermions with attractive interactions and 
two spin states will not collapse as bosanovae and supernovae, not
even near a Feshbach resonance because the kinetic energy dominates,
$1+\beta\ge0$.  Three spin states is marginal and may be studied in
$^6$Li where three broad Feshbach resonances lie close in magnetic field.
If three or more spin states can be tuned resonantly,
their attractive interaction energy should dominate as $a\to-\infty$ and
the systems collapse as ``Ferminovae''.

\item
Optical lattices in current experiments have few atoms in each local
trap and we thus expect superpairing which favors the insulator
vs. conductor state.

\end{itemize}

Tabletop experiments at low temperatures at Feshbach resonances extend
our studies of dilute systems to include also strongly interacting and
dense Fermi and Bose systems with unitarity scaling.  It will provide
new insight into a new scaling region as well as BCS pairing in atomic
traps. At certain interaction strengths and particle number the
pairing mechanisms and superfluidity in h.o. traps with cold atoms are
similar to that in nuclei and neutron star matter.

\end{document}